\documentclass[12pt,english]{article}
\usepackage[]{fontenc}
\usepackage[latin1]{inputenc}
\usepackage{amsmath}
\usepackage{amssymb}

\makeatletter

\newcommand{\noun}[1]{\textsc{#1}}

\usepackage{graphicx}
\usepackage{amssymb}
\usepackage{epstopdf}
\usepackage{amsmath}

\newcommand{\dslash}{{\not{{\hspace{-0.25em}}}} {\partial}}
\newcommand{\aslash}{{\not{{\hspace{-0.25em}}}} {\hspace{-0.25em}}A}

\newcommand{\jpi}{J_{{\pi}}}
\newcommand{\jsigma}{J_{{\sigma}}}
\newcommand{\W}{\overline{\mathcal{W}}}

\newcommand{\trc}{\rm{Tr_c}}
\newcommand{\Tint}{\int_0^{{\infty}} {\frac{dT}{T}} \exp  \left\{
-{\frac{m^2}{2}}T \right\}}
\newcommand{\pathint}{\int_{x,{\psi}} \exp  \{ -S_0 \}  \W}
\newcommand{\vsigma}{V_{{\sigma}}}
\newcommand{\vpi}{V_{{\pi}}}
\newcommand{\vzero}{V_0}

\usepackage{babel}

\usepackage{babel}
\makeatother
\begin{document}

\title{Super Wilson Loops in Planar QCD}

\date{Vikram Vyas%
\thanks{Associate, Abdus Salam International Centre for Theoretical Physics,
Trieste, Italy.%
} %
\thanks{Email: vikram@ajitfoundation.org%
}\\
 \textit{St. Stephen's College}\\
 \textit{Delhi University, Delhi 110 007, India}\\
 }

\maketitle
\begin{abstract}
In the planar limit of QCD meson correlation functions can be written
as a path-integral for a spin-half particle with each path being weighted
by the expectation value of the corresponding Super Wilson Loop. An
important quantity in this context is the expectation value of the
Super Wilson Loop averaged over loops of fixed length. I obtain the
leading and the sub-leading length dependence for this quantity. The
leading term, which was also known from the work of Banks and Casher,
reflects the fact that chiral symmetry is spontaneously broken in
planer QCD, while the sub-leading term implies that at least a finite
fraction of paths contributing to the average of the Super Wilson
Loop are effectively two-dimensional, thus suggesting a dual string
description of planar QCD.
\end{abstract}

\section{Introduction}

\label{sec:1} The planar limit of QCD, namely the limit in which
the number of colors $N$ goes to infinity, describes a theory of
free and stable mesons and glueballs and could be an appropriate starting
point for obtaining hadronic physics from QCD using a $1/N$ expansion~\cite{t'Hooft,witteN}.
This of course has been the main motivation behind the various efforts
towards solving planar QCD. Although none of these efforts have been
successful, there are features of planar QCD that continue to inspire
hope that it can indeed be solved. One of them is the fact that in
the planar limit the expectation value of the gauge-invariant observables
are given not by the full functional integral over all possible gauge
configurations, as one would expect in any quantum field theory, but
by its value on a single field, the so called ``master-field''~\cite{wittenCargese}.
Then there is an attractive possibility which is suggested by strong
coupling expansion in lattice gauge theory and by the recent developments
in string theory, that this master-field is given by a free string
theory~\cite{polyBook,adsCft}.

The gauge-invariant observable that best illuminates the possibility
of a dual string description of planar QCD are the various Wilson
loops~\cite{wilson74}. Thus in the planar limit we believe that
the expectation value of a Wilson loop should exhibit an area law
indicating that the planar QCD is a confining theory. Similarly, the
manner in which chiral symmetry is realized is reflected in the behavior
of the Wilson loop for a spin-half particle, often referred to as
the Super Wilson Loop (SWL). An important quantity for us will be
the expectation value of SWL averaged over loops of length $T$, which
we will denote by $<\W>_{T}$. The fact that chiral symmetry is spontaneously
broken in the planar limit~\cite{wittenColeman} implies that $<\W>_{T}$
must exhibit a leading behavior of the form $1/\sqrt{T}$~\cite{banksCasher}.
Once we recall that in $D$ dimensions the probability for a particle
to return to the origin after traveling a closed path of length $T$
is proportional to $1/T^{D/2}$, then this leading behavior implies
that at least a finite fraction of the paths contributing to the average
of a SWL are effectively one-dimensional. 

In fact the importance of the SWL in planar QCD is quiet general,
arising from the fact that the amplitude for a spin-half particle
to follow a given closed path in the presence of an external gauge
field is given, apart from the kinematic factors, precisely by the
corresponding SWL. This allows us to write meson propagators in planar
QCD as a sum over closed paths of a quark with each path being weighted
by the expectation value of the associated SWL. An efficient way of
writing such path-integrals is to use the worldline techniques~\cite{schuertRep}.

In this work we would like to further explore the behavior of the
SWL in the planar limit and relate it to the spectrum of the theory.
Specifically we will find that the leading $1/\sqrt{T}$ behavior
of the $<\W>_{T}$ will ensure the existence of massless pions, and
the assumption that the only massless spin-zero mesons are the Nambu-Goldstone
bosons will imply that the sub-leading behavior of a SWL must go as
$1/T^{n}$, where $n$ is greater than or equal to one. We will also
find that the difference between the scalar meson correlation function
and the pseudo-scalar meson correlation function crucially depends
on the sub-leading behaviour of $<\W>_{T}$, and that only sub-leading
behaviour of the form $1/T$ is consistant with the known short-distance
behaviour of these correlation functions. This sub-leading behaviour
implies that a fraction of paths that contribute to $<\W>_{T}$ are
effectively two-dimensional, and therefore is suggestive of a string
description of planar QCD.

The outline of this paper is as follows: In the next section I will
write down the connected two-point functions for spin-zero mesons
as a path integral for a spin-half particle using the so called vertex
operators. These operators play the role of interpolating fields in
the worldline representation. In section~\ref{sec:chiralSymmetry}
I will use these operators to relate the behavior of the SWL to the
spontaneous breaking of chiral symmetry and in this manner re-derive
the leading behaviour for $<\W>_{T}$. The main results are presented
in section~\ref{sec:results} where I obtain a relationship between
scalar and pseudo-scalar two point functions and use it to obtain
the sub-leading dependence of $<\W>_{T}$ on $T$. In the final section
I discuss these results in the context of a possible string description
of the planar QCD.

\section{\label{sec:vertexOperators}Vertex Operators for Spin-Zero Mesons}

To motivate the use of worldline representation let us first recall
one of the simplifying features of planar QCD. Consider the Euclidean
partition function of a $SU(N)$ gauge theory with one flavor of quark%
\footnote{In the planar limit the flavor degrees of freedom will not play any
dynamical role and there is a genuine Nambu-Goldstone boson corresponding
to the spontaneous breaking of $U(1)$ flavour symmetry.%
} of mass $m$ in the presence of the sources for the scalar and the
pseudoscalar mesons, \begin{eqnarray}
Z[J] & = & \int_{A}\exp\left(-S[A]\right)\exp\left(-\Gamma[A,J]\right),\label{Z}\\
-\Gamma[A,J] & = & \ln\det(-i\dslash-\aslash-im-i\jsigma-\gamma_{5}\jpi),\label{fermDet}\end{eqnarray}
 where $S[A]$ is the Yang-Mill action for the $SU(N)$ gauge field
$A$. This can be written as\begin{equation}
Z[J]=Z[0]\left<\exp\left(-\Gamma[A,J]\right)\right>_{YM},\end{equation}
 where the functional average is with respect to the Yang-Mill action
$S[A]$. In the planar limit, as we have noted in the introduction,
only one gauge field configuration, the so called master field $\bar{A}$,
contributes to the functional integral~\cite{witteN}. Therefore,
the partition function can be written as \begin{equation}
Z[J]=Z[0]\exp\left(-\Gamma[\bar{A},J]\right).\label{planarZ}\end{equation}
 Once we know $Z[J]$ we can of course obtain various connected two-point
functions by differentiating $\ln Z[J]$ with respect to suitable
sources. Thus in the planar limit the scalar and pseudo-scalar two-point
functions,

\begin{eqnarray}
\Delta_{s}(X-Y) & = & <\Omega|\bar{\Psi}\Psi(X)\bar{\Psi}\Psi(Y)|\Omega>,\\
\Delta_{p}(X-Y) & = & <\Omega|\bar{\Psi}i\gamma_{5}\Psi(X)\bar{\Psi}i\gamma_{5}\Psi(Y)|\Omega>,\end{eqnarray}
 can be written as

\begin{eqnarray}
\Delta_{p}(X-Y) & = & \left(\frac{\delta}{\delta\jpi(X)}\frac{\delta}{\delta\jpi(Y)}\Gamma[\bar{A},J]\right)_{J=0},\label{pionProp}\\
\Delta_{s}(X-Y) & = & \left(\frac{\delta}{\delta\jsigma(X)}\frac{\delta}{\delta\jsigma(Y)}\Gamma[\bar{A},J]\right)_{J=0}.\label{sigmaProp}\end{eqnarray}
Now the real part of the fermionic effective action, $\Gamma[\bar{A},J]$,
which is what is required for the above two-point functions, can be
written as a path-integral for a spin-half particle~\cite{schubert1,dHoker1}.
\begin{eqnarray}
\Gamma[\bar{A},J] & = & \int_{0}^{\infty}\frac{dT}{T}\exp\{-m^{2}\frac{T}{2}\}\label{WLeffA}\\
 & \times & \int_{x,\psi_{a}}\exp\{-(S_{0}+S[J])\}\W,\nonumber \end{eqnarray}
 where the worldline of a particle is specified by four bosonic coordinates
$x_{\mu}(\tau)$ which satisfy periodic boundary conditions and by
six fermionic, or the anti-commuting, coordinates $\psi_{a}(\tau)$
that satisfy anti-periodic boundary conditions (see Ref.~\cite{dHoker1}
for details.) The worldline actions $S_{0}$ and $S[J]$ are given
by \begin{eqnarray}
S_{0} & = & \int_{0}^{T}d\tau\Big\{\frac{\dot{x}^{2}}{2}+\frac{1}{2}\psi_{\mu}\dot{\psi_{\mu}}+\frac{1}{2}\psi_{5}\dot{\psi_{5}}+\frac{1}{2}\psi_{6}\dot{\psi_{6}}\Big\}\label{eq:s0}\\
S[J] & = & \int_{0}^{T}d\tau\Big\{\frac{1}{2}J_{\pi}^{2}+i\psi_{\mu}\psi_{5}\partial_{\mu}J_{\pi}+mJ_{\sigma}+\frac{1}{2}J_{\sigma}^{2}+i\psi_{\mu}\psi_{6}\partial_{\mu}J_{\sigma}\Big\}\label{eq:sj}\end{eqnarray}
 while the expectation value of the SWL, $\W$, is defined as\begin{eqnarray}
\W & = & \left<\trc\hat{P}\exp\left\{ i\int_{0}^{T}d\tau\{\dot{x}_{\mu}A_{\mu}-\frac{1}{2}\psi_{\mu}F_{\mu\nu}\psi_{\nu}\right\} \right>_{YM}\label{fwl}\\
 & = & \trc\hat{P}\exp\left\{ i\int_{0}^{T}d\tau\{\dot{x}_{\mu}\bar{A}_{\mu}-\frac{1}{2}\psi_{\mu}\bar{F}_{\mu\nu}\psi_{\nu}\right\} \end{eqnarray}
 where $F_{\mu\nu}$ is the Yang-Mills field strength tensor while
the trace is over the color degrees of freedom.

Using this path-integral representation of the one loop fermionic
effective action in (\ref{pionProp}) and (\ref{sigmaProp}) we can
readily obtain the worldline representation for the scalar and the
pseudoscalar two-point functions. For this purpose it is convenient
to define a worldline average of any functional $F[x,\psi]$ as \begin{equation}
<F>_{wl}=\Tint\pathint F\label{eq:wlaverage}\end{equation}
 then the two-point functions can be written as\begin{eqnarray}
\Delta_{p}(X-Y) & = & <\vpi(X)\vpi(Y)>_{wl},\label{eq:wlinePion}\\
\Delta_{s}(X-Y) & = & <\vsigma(X)\vsigma(Y)>_{wl},\label{eq:wlineSigma}\end{eqnarray}
 where the pseudo-scalar vertex operator $\vpi(X)$ is given by \begin{eqnarray}
\vpi(X) & = & i\int_{0}^{T}d\tau\psi_{\mu}\psi_{5}\partial_{\mu}\delta(x(\tau)-X),\end{eqnarray}
 while the scalar vertex operator $V_{\sigma}(X)$ is \begin{eqnarray}
\vsigma(X) & = & \int_{0}^{T}d\tau\Big\{ m\delta(x(\tau)-X)\label{Vsigma}\\
 & + & i\psi_{\mu}(\tau)\psi_{6}(\tau)\partial_{\mu}\delta(x(\tau)-X)\Big\}.\nonumber \end{eqnarray}
 In Eqs (\ref{eq:wlinePion}) and (\ref{eq:wlineSigma}) a contact
term which will not play any role in the present investigation has
been neglected%
\footnote{The contact term is of the following form $\int_{0}^{T}d\tau\delta(X-x(\tau))\delta(Y-x(\tau))$. %
}. It will be convenient to write the scalar vertex operator as \begin{equation}
\vsigma(X)=m\vzero(X)+V_{s}(X),\label{Vscalar}\end{equation}
 where \begin{eqnarray}
\vzero(X) & = & \int_{0}^{T}d\tau\delta(x(\tau)-X),\label{Vzero}\\
V_{s}(X) & = & i\int_{0}^{T}d\tau\psi_{\mu}(\tau)\psi_{6}(\tau)\partial_{\mu}\delta(x(\tau)-X).\label{Vsix}\end{eqnarray}
It is useful to observe that formally the above vertex operators can
be defined as\begin{eqnarray}
V_{\pi}(X) & = & \left(\frac{\delta}{\delta J_{\pi}(X)}S[J]\right)_{J=0}\label{eq:Vpi}\\
V_{\sigma}(X) & = & \left(\frac{\delta}{\delta J_{\sigma}(X)}S[J]\right)_{J=0},\label{eq:Vsigma}\end{eqnarray}
where $S[J]$ is given by (\ref{eq:sj}). 

Finally, we note that the left hand side of (\ref{eq:wlinePion})
and (\ref{eq:wlineSigma}) represents a formal sum of all the planar
diagrams which have quark line only at the boundary of the diagram.
Since in the planar limit these are the only diagrams that contribute
to the meson correlation function we could have written these equations
without any reference to a master-field.

\section{Chiral Condensate and the SWL}

\label{sec:chiralSymmetry} In the planar limit the expectation value
of the chiral condensate can be written as the worldline average of
the scalar vertex operator $V_{\sigma}$ \begin{equation}
<\Omega|\bar{\Psi}\Psi(X)|\Omega>=i<V_{\sigma}(X)>_{wl}.\end{equation}
 Denoting the worldline average of \textbf{$V_{\sigma}$} by $V_{\chi}$
and using  (\ref{Vscalar}) we obtain\begin{eqnarray*}
V_{\chi} & = & m<V_{0}(X)>_{wl}+<V_{s}(X)>_{wl}\\
 & = & m<V_{0}(X)>_{wl}\end{eqnarray*}
where we have used the fact that $<V_{s}(X)>_{wl}$ vanishes as it
is odd in the Grassmann variable $\psi_{6}$. Using our definition
of the worldline average, (\ref{eq:wlaverage}), we obtain the path-integral
representation for the chiral condensate,\begin{equation}
V_{\chi}=m\int_{0}^{{\infty}}\frac{dT}{T}\exp\left\{ -{\frac{m^{2}}{2}}T\right\} \pathint\int_{0}^{T}d\tau\delta(x(\tau)-X).\end{equation}
 In performing the path-integral over $x(\tau)$ it is convenient
to separate the zero mode which arises because of the translation
invariance. One way of doing this is to write the path $x(\tau)$
as\begin{equation}
x(\tau)=\bar{x}+y(\tau),\label{eq:zeroMode}\end{equation}
where\[
\bar{x}=\frac{1}{T}\int_{0}^{T}d\tau x(\tau),\]
\[
\int_{0}^{T}d\tau y(\tau)=0.\]
 The path-integral then can be decomposed into an integral over $\bar{x}$
and a path-integral over $y(\tau)$, \begin{equation}
V_{\chi}=m\Tint\int_{y,\psi}d\bar{x}\exp\{-S_{0}\}\W\int_{0}^{T}d\tau\delta(\bar{x}+y(\tau)-X),\label{eq:xbaryPInt}\end{equation}
 writting the delta function in the momentum representation and then
doing the integral over $\bar{x}$ leads to\begin{equation}
V_{\chi}=m\int_{0}^{\infty}dT\exp\left\{ -\frac{m^{2}}{2}T\right\} \int_{y,\psi}\exp\{-S_{0}\}\W.\label{eq:yPIforChi}\end{equation}
 This is essentially the expression for chiral-condensate that Banks
and Casher had obtained in~\cite{banksCasher}, except that instead
of trace over gamma-matrices we have an integral over the Grassmann
variables $\psi$. 

The spontaneous breaking of chiral symmetry requires that $V_{\chi}$
does not vanish in the limit of the current quark mass $m$ going
to zero. This is possible if the worldline average of the SWL of length
$T$ has the following asymptotic behaviour\begin{eqnarray}
<\W>_{T} & \equiv & \lim_{T\rightarrow\infty}\int_{y,\psi}\exp\{-S_{0}\}\W\nonumber \\
 & = & \frac{C_{0}}{T^{1/2}}+\mathcal{O}\left(\frac{1}{T^{n}}\right),\label{eq:banksCasher}\end{eqnarray}
 where $C_{0}$ is some constant and $n>\frac{1}{2}$. It is interesting
to note that in $D$ dim\begin{equation}
\lim_{T\rightarrow\infty}\int_{y,\psi}\exp\left\{ -S_{0}\right\} \propto\frac{1}{T^{D/2}},\label{eq:randomWalk}\end{equation}
 this can be thought of as the probability for a free particle to
return to its starting point after following a closed path of length
$T$. The same probability in the presence of external gauge fields
is given by (\ref{eq:banksCasher}), thus the effect of $\W$ is to
increase the probability for the existence of a closed path of length
$T$ from its value in four dimension to that of one-dimension. Correspondingly,
in the Minkowski space-time the effect of $\W$ is to increase the
amplitude for process in which a quark-antiquark pair is created and
lasts for a (proper) time of the order of $T$ from its value in four-dimension
to that of its value in one-dimension. As emphasized in~\cite{banksCasher},
it is the spin-dependent interaction, described by the SWL, which
is responsible for the existence of effectively one-dimensional paths
that ensures that the leading behavior of $<\W>_{T}$ is proportional
to $1/T^{1/2}$.

\section{Chiral Limit of Planar QCD\label{sec:results}}

In the previous section we found the asymptotic behaviour of the SWL
which is responsible for the spontaneous breaking of chiral symmetry.
Now we would like to see how this asymptotic behaviour influences
the spectrum of the spin-zero mesons. Consider first the scalar two-point
function~(\ref{sigmaProp}) which can be written in the worldline
representation as \begin{equation}
\Delta_{s}(k)=<V_{s}(k)V_{s}(-k)>_{wl}+m^{2}<V_{0}(k)V_{0}(-k)>_{wl},\label{wlineSigmaProp}\end{equation}
 where the vertex function $V_{s}$ and $V_{0}$ are given by~(\ref{Vsix})
and (\ref{Vzero}) and we have used the fact that the cross term $<V_{0}V_{s}>_{wl}$
vanishes being odd in the Grassmann variable $\psi_{6}$. Next we
notice that \begin{equation}
<\vpi(k)\vpi(-k)>_{wl}=<V_{s}(k)V_{s}(-k)>_{wl}\label{V5equalsV6}\end{equation}
 as the right hand side can be obtained from the left hand side by
simply relabeling $\psi_{6}$ as $\psi_{5}$ and $\psi_{5}$ as $\psi_{6}$
in (\ref{eq:wlaverage}). Using this in~(\ref{wlineSigmaProp}) we
obtain the following relation between the scalar and the pseudo-scalar
two-point function \begin{equation}
\Delta_{p}(k)=\Delta_{s}(k)-m^{2}<V_{0}(k)V_{0}(-k)>_{wl},\label{chiralWId1}\end{equation}
which can be written in a more suggestive form as\begin{equation}
\Delta_{p}(k)-\Delta_{s}(k)=-m^{2}<V_{0}(k)V_{0}(-k)>_{wl}.\label{eq:chiralWId}\end{equation}
Consider this relation in the chiral limit, if\begin{equation}
\lim_{m\rightarrow0}<V_{0}(k)V_{0}(-k)>_{wl}=0\label{eq:symmetricPhase}\end{equation}
 then we would have \begin{equation}
\Delta_{p}(k)=\Delta_{s}(k),\label{eq:symmetricPropagators}\end{equation}
which would correspond to the situation when chiral symmetry is not
spontaneously broken. In the planar limit we know that chiral symmetry
is indeed spontaneously broken~\cite{wittenColeman} therefore \begin{equation}
\lim_{m\rightarrow0}m^{2}<V_{0}(k)V_{0}(-k)>_{wl}\neq0.\label{chiralOrderParameter}\end{equation}

We also know that planar QCD is a theory of an infinite number of
free, stable, and non-interacting, mesons and glueballs~\cite{witteN}.
This in turn implies that the meson correlation functions are of the
form of an infinite sum of simple poles. In particular the scalar
meson correlation function $\Delta_{s}(k)$ has the following form
\begin{eqnarray}
\Delta_{s}(k) & = & \sum_{n=1}^{\infty}\frac{Z_{s,n}^{2}}{k^{2}+M_{s,n}^{2}}.\label{largeNsTwoPoint}\end{eqnarray}
 where $Z_{n,s}=<\Omega|\bar{\Psi}\Psi|n>$ is the matrix element
for $\bar{\Psi}\Psi$ to create $n^{th}$ scalar meson from the vacuum.
Similarly the pseudo-scalar correlation function can be written as\begin{equation}
\Delta_{p}(k)=\frac{Z_{\pi}^{2}}{k^{2}+M_{\pi}^{2}}+\sum_{n=1}^{\infty}\frac{Z_{p,n}^{2}}{k^{2}+M_{p,n}^{2}},\label{chiralWId2}\end{equation}
 where we have separated the pion pole and the pion mass vanishes
in the chiral limit as\[
\lim_{m\rightarrow0}M_{\pi}^{2}=C_{\chi}m,\]
where $C_{\chi}$ is related to the chiral condensate. Combining (\ref{eq:chiralWId})
with (\ref{largeNsTwoPoint}) and (\ref{chiralWId2}) leads to\begin{eqnarray*}
\Delta_{P}(k)-\Delta_{s}(k) & = & -m^{2}<V_{0}(k)V_{0}(-k)>_{wl}\\
 & = & \frac{Z_{\pi}^{2}}{k^{2}+M_{\pi}^{2}}+\sum_{n=1}^{\infty}\frac{Z_{p,n}^{2}}{k^{2}+M_{p,n}^{2}}-\sum_{n=1}^{\infty}\frac{Z_{s,n}^{2}}{k^{2}+M_{s,n}^{2}}\end{eqnarray*}
 Consider the above expression in the $\lim k\rightarrow0$ with a
small but finite current quark mass \begin{eqnarray}
\lim_{k\rightarrow0}-m^{2}<V_{0}(k)V_{0}(-k)>_{wl} & = & \frac{1}{m}\frac{Z_{\pi}^{2}}{C_{\chi}}+\sum_{n=1}^{\infty}\frac{Z_{p,n}^{2}}{M_{p,n}^{2}}-\sum_{n=1}^{\infty}\frac{Z_{s,n}^{2}}{M_{s,n}^{2}}.\label{genreralVVkGRO}\end{eqnarray}
 Now in the limit of $k\rightarrow0$ we can evaluate the left hand
side of the above expression \begin{multline}
\lim_{k\rightarrow0}-m^{2}<V_{0}(k)V_{0}(-k)>_{wl}=-m^{2}\int_{0}^{\infty}dTT\exp\left\{ -{\frac{m^{2}}{2}}T\right\} \\
\times\int_{y,\psi}\exp\{-S_{0}\}\W.\label{pionmass}\end{multline}
 where I have used \begin{equation}
V_{0}(k)=\int_{0}^{T}d\tau\exp\{\imath k\cdot y(\tau)\}.\label{V0k}\end{equation}
 As noted in Eq.(\ref{eq:banksCasher}), if chiral symmetry is spontaneously
broken then we must have the following asymptotic behaviour \begin{eqnarray}
\lim_{T\rightarrow\infty}<\W>_{T} & \equiv & \lim_{T\rightarrow\infty}\int_{y,\psi}\exp\{-S_{0}\}\W\label{eq:defTavg}\\
 & = & \frac{C_{0}}{\sqrt{T}}+\frac{C_{1}}{T^{n}}+\cdots\label{eq:asympSWL}\end{eqnarray}
with some constants $C_{0}$ and $C_{1}$. The second term on the
right hand side with $n>\frac{1}{2}$ represents the next to the leading
term in the asymptotic expansion. Using this in~(\ref{pionmass})
leads to \begin{equation}
\lim_{k\rightarrow0}-m^{2}<V_{0}(k)V_{0}(-k)>_{wl}=\frac{\tilde{C_{0}}}{m}+\tilde{C_{1}}m^{2n-2},\label{wlresult}\end{equation}
 with constants $\tilde{C_{0}}$ and $\tilde{C_{1}}$ related to $C_{0}$
and $C_{1}$ by known factors. Comparing the above result with (\ref{genreralVVkGRO})
we find that the first term on the right hand side of the above expression
arises from the pion pole with $M_{\pi}^{2}$ proportional to the
current quark mass $m$, while the second term implies that $n\geq1$,
assuming that in the chiral limit only massless spin-zero particle
are the Nambu-Goldstone bosons.

From (\ref{wlresult}) we also see that the difference between the
pseudo-scalar and the scalar two-point function  crucially depends
on whether $n=1$ or $n>1$. Consider first the possibility that $n>1$.
If that is the case, in the chiral limit (\ref{wlresult}) is consistent
only with \begin{eqnarray}
\lim_{m\rightarrow0}\Delta_{\pi}(k)-\Delta_{\sigma}(k) & = & -m^{2}<V_{0}(k)V_{0}(-k)>_{wl}\nonumber \\
 & = & \frac{Z_{\pi}^{2}}{k^{2}+M_{\pi}^{2}},\label{eq:nEqualOne}\end{eqnarray}
with $M_{\pi}^{2}$ vanishing linearly with the current quark mass
$m$, but we know the momentum dependence of the right hand side of
the above equation for large momentum from \cite{shifmanOPE} and
it goes as $1/k^{4}$ in the chiral limit. Therefore, the possibility
of $n>1$ is excluded and the long distance behaviour of the expectation
value of SWL averaged over paths of length $T$ is given by\begin{equation}
\lim_{T\rightarrow\infty}<\W>_{T}=\frac{C_{0}}{T^{1/2}}+\frac{C_{1}}{T}+...\label{eq:mainResult}\end{equation}
 This result has an interesting geometrical interpretation, the meaning
of the leading term was already discussed in section (\ref{sec:chiralSymmetry}).
Let us consider now the sub-leading term, recalling again that in
$D$ dimensions the probability of a particle to return to the origin
after transversing a path of length $T$ goes as $T^{-\frac{D}{2}}$,
therefore the sub-leading behaviour of the form of $T^{-1}$ implies
that a finite fraction of paths that contribute to $<\W>_{T}$ are
effectively two-dimensional. 

Let us consider (\ref{eq:mainResult}) in the context of a possible
dual string description of the planar QCD. In such a description mesons
would appear as open strings with quark or an anti-quark as the end
point of the string. The leading term in $<\W>_{T}$ then suggests
that there must be a spin-string interaction between quarks and the
string connecting them. This spin-string interaction leads to an effective
short range spin-spin interaction which is responsible for the $1/T^{1/2}$
behaviour of the leading term. Now consider the sub-leading term.
This term is quite natural in a string description of mesons, as it
implies that at least a finite fraction of paths that contribute to
the average of the expectation value of the SWL are effectively two-dimensional.

\section{Conclusions\label{sec:Conclusions}}

There is a persistent hope that planar QCD can be solved by reformulating
it as a string theory. In the planar limit the observables that seems
most amenable to a string representation are the various types of
Wilson loops. For planar QCD with quarks the appropriate Wilson loop
in terms of which one can write the various meson Greens functions
is the Wilson loop associated with a spin-half particle or the SWL.
As we have seen in previous sections, an important quantity that probes
the nature of the meson spectrum in the planar limit is the expectation
value of the SWL averaged over loops of fixed length $T$, which we
have denoted by $<\W>_{T}$. 

In the present work we have been able to determine the leading and
the subleading $T$ dependence for this quantity. They in turn suggest
a string model of planar QCD in which there is a spin-string interaction
between the spin of the quarks and the string connecting them. The
spin-string interaction leads to an effective spin-spin interaction
between the spin of the quark and the spin of the anti-quark which
is responsible for the leading behavior of $<\W>_{T}$ and is responsible
for the spontaneous breaking of chiral symmetry. The sub-leading term
is more directly connected with the string behaviour as it suggests
that at least a finite fraction of the paths contributing to $<\W>_{T}$
are effectively two-dimensional.

The present analysis therefore implies that in seeking a dual string
representation of planar QCD one must include spin-string interaction
in order to correctly reflect the manner in which chiral symmetry
is realised. One approach, which has been investigated in~\cite{vv00},
is to seek a string representation for the expectation value of the
SWL. For SWL in a compact $U(1)$ gauge theory, which is known to
be confining~\cite{polyCstring}, one indeed finds that the strings
describing the SWL have spin-string interaction which leads to an
effective spin-spin interaction between the spin of the quark and
the anti-quark. Of course the more important and interesting challenge
is to find a string representation for the SWL in planar QCD which
correctly incorporates the spin-string interaction.

\section*{Acknowledgment}

I have benefited immensely from my discussion with K. S. Narain and
would like to thank him for that. I would also like to thank Ansar
Fayyazuddin, Gunnar Bali and M. Shifman for their critical comments
at various stages of this investigation. A large part of this work
was done while I was visiting the Abdus Salam International Centre
for Theoretical Physics, Trieste. I would like to thank the Associate
program of the ICTP for making this visit possible.

\end{document}